\documentclass[aip,superscriptaddress,amsmath,amssymb,floatfix,reprint]{revtex4-1}
\pdfoutput=1
\usepackage{amssymb}
\usepackage{epstopdf}
\usepackage{graphicx}
\usepackage{bm}
\usepackage[pdftex,breaklinks=true,bookmarksopen=true,bookmarksopenlevel=3,bookmarksnumbered=true,colorlinks=true,urlcolor= magenta,citecolor=blue,linkcolor=blue]{hyperref}

\begin{document}

\title{Observation of longitudinal and transverse self-injections in laser-plasma accelerators}

\author{S. Corde}
\thanks{The first two authors have contributed equally.}
\affiliation{Laboratoire d'Optique Appliqu\'ee, ENSTA ParisTech - CNRS UMR7639 - \'Ecole Polytechnique, 828 Boulevard des Mar\'echaux, 91762 Palaiseau, France}
\author{C. Thaury}
\thanks{The first two authors have contributed equally.}
\affiliation{Laboratoire d'Optique Appliqu\'ee, ENSTA ParisTech - CNRS UMR7639 - \'Ecole Polytechnique, 828 Boulevard des Mar\'echaux, 91762 Palaiseau, France}
\author{A. Lifschitz}
\author{G. Lambert}
\author{K. Ta Phuoc}
\affiliation{Laboratoire d'Optique Appliqu\'ee, ENSTA ParisTech - CNRS UMR7639 - \'Ecole Polytechnique, 828 Boulevard des Mar\'echaux, 91762 Palaiseau, France}
\author{X. Davoine}
\affiliation{CEA, DAM, DIF, F-91297 Arpajon, France}
\author{R. Lehe}
\author{D. Douillet}
\author{A. Rousse}
\author{V. Malka}
\affiliation{Laboratoire d'Optique Appliqu\'ee, ENSTA ParisTech - CNRS UMR7639 - \'Ecole Polytechnique, 828 Boulevard des Mar\'echaux, 91762 Palaiseau, France}

\begin{abstract}
Laser-plasma accelerators can produce high quality electron beams, up to giga-electronvolts in energy,  from a centimeter scale device. The properties of the electron beams and the accelerator stability are largely determined by the injection stage of electrons into the accelerator. The simplest mechanism of injection is self-injection, in which the wakefield is strong enough to trap cold plasma electrons into the laser wake. The main drawback of this method is its lack of shot-to-shot stability. Here we present experimental and numerical results that demonstrate the existence of two different self-injection mechanisms. Transverse self-injection is shown to lead to low stability and poor quality electron beams, because of a strong dependence on the intensity profile of the laser pulse. In contrast, longitudinal injection, which is unambiguously observed  for the first time,  is shown to lead to  much more stable acceleration and higher quality electron beams. 
\end{abstract}

\maketitle 

\noindent
Laser-plasma accelerators~\cite{PRL1979Tajima,RMP2009Esarey} have gained considerable attention for their potential in delivering, with a compact setup, quasi-monoenergetic~\cite{Nature2004Faure,Nature2004Geddes,Nature2004mangles}, low emittance~\cite{PRL2010Brunetti,2012PRL_Plateau,weingartner2012}, giga-electronvolt electron beams~\cite{NatPhys2006Leemans}. The few femtosecond duration of the accelerated electrons bunches~\cite{NatPhys2011Lundh}, associated to a peak current of a few kiloamps, make these accelerators of particular interest for applications~\cite{NatPhys2008Malka}, such as studies of electron-induced damage in biological tissues at very high dose rate~\cite{CDD2011Rigaud}, or short wavelength, femtosecond and micrometer source size light beams~\cite{PRL2004Rousse, NatPhys2010Kneip,NatPhys2008Nakajima,NatPhys2008Schlenvoigt,NatPhys2009Fuchs,NatPhys2011Klopper,kim_nat_2012}.    
The development of these accelerators has however been hindered by a lack of stability of the acceleration and a poor control on the electron features. These problems can be largely traced back to the injection of electrons in the plasma wake driven by the laser pulse. Electron trapping is generally achieved by the wave-breaking of the plasma wake, a process that is by nature uncontrollable and leads generally to poor quality electrons.  Controlled injection techniques, \emph{e.g.}, colliding pulse injection~\cite{Nature2006Faure}, ionization-induced injection~\cite{PRL2011Pollock} and density gradient injection~\cite{NatPhys2011Gonsalves}, have been developed to overcome this shortcoming. 

\begin{figure}[b]
\includegraphics[width=9cm]{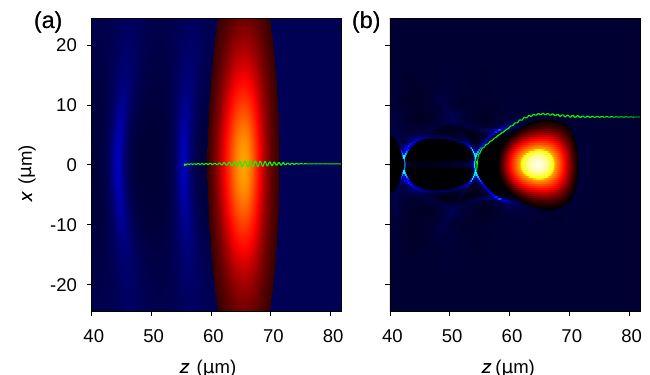}
\caption{\textbf{Schematic for longitudinal and transverse self-injections.} (a) Typical trajectory of an injected electron in the longitudinal self-injection mechanism.  (b) Typical trajectory of an injected electron in the transverse self-injection mechanism. The blue colour scale represents the electron density. The red to yelow colour scale indicates the laser intensity. The trajectories are given by the green lines.}
\label{fig:fig1}
\end{figure}

These methods offer an improved control on the acceleration and lead to better electron features, but they imply generally complex setups. For this reason, self-injection remains the most common method for injecting electrons in the plasma wake. Two distinct physical mechanisms can be distinguished~: longitudinal and transverse self-injection. In longitudinal self-injection, the trajectory of injected electrons is mainly longitudinal, with a negligible transverse motion. As shown by the schematic in Fig.~1a, the injected electrons pass through the laser pulse and gain energy while crossing the plasma wave. When they reach the rear of the first plasma period, their velocity exceeds the wake phase velocity and the electrons are eventually injected. The only electrons that are trapped are those that were initially close to the axis ($r\approx0$) where the laser intensity and the wakefield amplitude are the highest and where the ponderomotive force is small. The longitudinal self-injection mechanism is analogous to one-dimensional longitudinal wave-breaking~\cite{SovPhysJETP1956Akhiezer}. In contrast, transverse self-injection \cite{PRL1997Bulanov} is a multi-dimensional effect. It occurs in the bubble regime~\cite{APB2002Pukhov,pop2004zhidkov,PRL2006Lu}, where the laser ponderomotive force expels electrons from the propagation axis and forms an electron-free cavity in its wake. As shown in Fig.~1b, the injected electrons are initially located at approximately one laser waist from the axis ($r\sim w_0$). They circulate around the laser pulse and the bubble, and attain a velocity larger than the wake phase velocity when  reaching the axis at the rear of the bubble~\cite{ PoP2006Tsung, PRL2009Kostyukov, PoP2009Wu}. Except in a few papers (\emph{e.g.} Ref.~\cite{PRL2010Brunetti}), trapping of quasi-monoenergetic electrons is generally attributed to transverse self-injection.

\begin{figure*}[t]
\includegraphics[width=14cm]{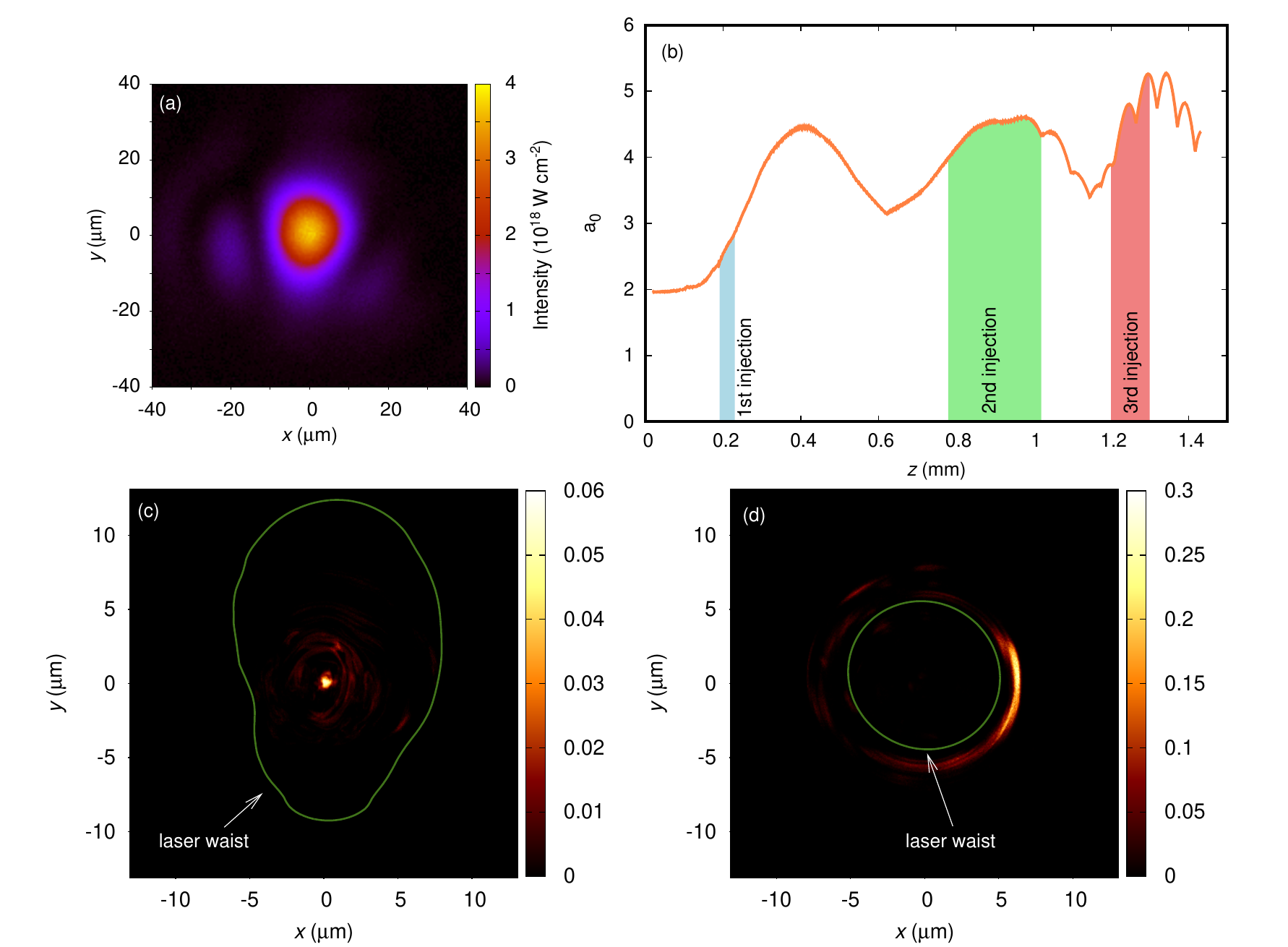}
\caption{\textbf{Particle-in-cell simulation of the interaction.} (a) Focal spot  intensity distribution, taken from experimental data. (b) Evolution  of the normalized laser amplitude. The coloured areas indicate the injection regions. (c-d) Distribution of the transverse positions from which electrons injected in the  first (c) and second bunches (d) originate (the density is in arbitrary units). The $1/e^2$ contours of the laser pulse head are indicated in violet. The laser waist is smaller in (d) than in (c) because of self-focusing.}
\label{fig:fig2}
\end{figure*}

This article presents experimental and simulation results which demonstrate that, depending on laser and plasma parameters, electrons can be trapped either via longitudinal or transverse self-injection. These two types of injection are associated with energetic electrons with different properties, which allows them  to be identified unambiguously. 
First, we show that in some conditions, electrons can be trapped successively by longitudinal followed by transverse self-injection. The experimental observation of this phenomenon was made possible by the use of a  gas cell with an adjustable length. Second, we prove that trapping via longitudinal self-injection provides a stable and tunable source of energetic electrons. 

\bigskip
\noindent
{\textbf{Results}

\smallskip
\noindent
\textbf{Self-injection mechanisms.}
Transverse self-injection is known from 3D particle-in-cell (PIC) simulations to be the most efficient trapping mechanism in the wake of an ultra-short laser pulse~\cite{APB2002Pukhov,PRL2006Lu, PoP2006Tsung,PRL2009Kostyukov}. We show here that this injection can, though, be preceded by a noticeable trapping via longitudinal self-injection.  We focus first on PIC simulations. As the aberrations on the laser beam are known to play a role on electron injection~\cite{glinec2008,apl2009mangles}, the transverse profile of laser intensity used in the simulation was taken from experimental data (Fig.~2a).  The laser has a elliptical shape with some contribution of high-order modes.

Figure~2b shows the evolution of the normalized laser amplitude $a_0$ for a 30 fs laser pulse, with an initial amplitude $a_0=2$, propagating in a plasma with density $n_0=10^{19}$ cm$^{-3}$ (see Methods for details on the simulation). The laser pulse undergoes three cycles of focusing-defocusing due to relativistic self-focusing and self-steepening, leading to multiple injection~\cite{CPL2008Yoshitima, PoP2008Oguchi}. A 2 pC bunch is injected very early in the propagation, before the first self-focusing peak, whereas a much more charged bunch (200 pC) is injected around the second self-focusing peak. A less charged bunch is injected around the third self-focusing peak. 

The initial spatial distributions in the transverse plane of the electrons which are eventually injected are shown in Figs.~2c and~2d for the first and the second bunches respectively. Electrons in the second bunch originate from positions close to the laser waist, as expected in the case of trapping by transverse self-injection~\cite{PRL2009Kostyukov}.  In contrast, electrons in the first bunch come from regions close to the axis.
When these electrons are injected, the laser spot radius is large and $a_0$ is still low ($a_0\approx2.5$), so the radial ponderomotive force close to the axis is small. Thus on-axis electrons are only weakly deviated when crossing the laser pulse, and they remain in the region of largest accelerating field $E_z$. 
Moreover, the  laser amplitude increases steeply in the region of first injection, due to laser self-focusing  (see Fig.~2b). This reduces the wake phase velocity via the relativistic-shift of the plasma wavelength $\lambda_p$~\cite{kalmykov_njp_2010}. The strongly reduced wake phase velocity lowers the threshold for trapping such that electrons can catch up the plasma wave and be injected despite a low $a_0$, similarly to density-gradient injection~\cite{NatPhys2011Gonsalves}.  

Figure~2b shows that the conditions required to trigger longitudinal injection are fulfilled  during $\approx 40$ $\mu$m only. In contrast, transverse injection occurs on $\approx 240$ $\mu$m. As a result, longitudinal injection leads to the acceleration of quasi-monoenergetic electron beams, whereas transverse injection generally results in the production of broadband electron spectra.   Figure~2 also indicates that longitudinal injection takes place at the very beginning of the interaction (after $\approx 200$ $\mu$m of laser propagation), while transverse injection is triggered after significant self-focusing of the laser.  As the charge in the second bunch is much larger than in the first bunch, only transverse injection is observed in experiments where the plasma length is typically in the range $2-10$ mm.

A second feature visible in Fig.~2d is the strong asymmetry on the initial spatial distribution of electrons in the second bunch, when transverse injection models predict an even distribution around the laser waist~\cite{PRL2009Kostyukov}. This asymmetry is due to the asymmetry on the laser intensity distribution close to the waist, which breaks the cylindrical symmetry of the plasma cavity. 
Small changes in the focal spot lead to significant variations of the transverse distribution. Transverse injection is, hence, intrinsically  unstable. 

To observe experimentally sequential longitudinal and transverse injections, we used the ``Salle jaune'' 30 TW laser system to accelerate electrons in a gas cell which length can be adjusted from 200 $\mu$m up to 10 mm (see Methods). 
Typical experimental electron spectra, obtained for different cell lengths $L$ and for an electron density $n_e\approx 1.1\times 10^{19}$~cm$^{-3}$, are plotted in Fig.~3a. For short lengths, $L\lesssim 1.4$~mm, spectra consist of quasi-monoenergetic beams with a charge in the $2-10$~pC range. For $L=1.4$~mm,  a second electron bunch, with a low energy, is observed. This second bunch is then accelerated to higher energy as the cell length is increased. It exhibits a broad spectrum and has a charge one order of magnitude higher than the first bunch, in the 50-100 pC range.
The mean cut-off energies measured for  the two electron bunches are shown in Fig.~3b. The low charge (first) bunch reaches a maximum energy of $\approx 250$~MeV for $L\approx 1$ mm. Then it decreases from $L=1.4$~mm, indicating that this bunch has outrun the plasma wave and entered a dephasing region where it is decelerated by the wakefield~\cite{PRL1979Tajima}. In contrast, the energy of the high charge bunch keeps increasing up to $\approx 270$ MeV for $L=5$~mm. As a result, the energy of the high charge bunch overtakes the one of the low charge, for $L=2.1$~mm, preventing the detection of the first bunch for long plasmas.

\begin{figure}
	\includegraphics[width=7.5cm]{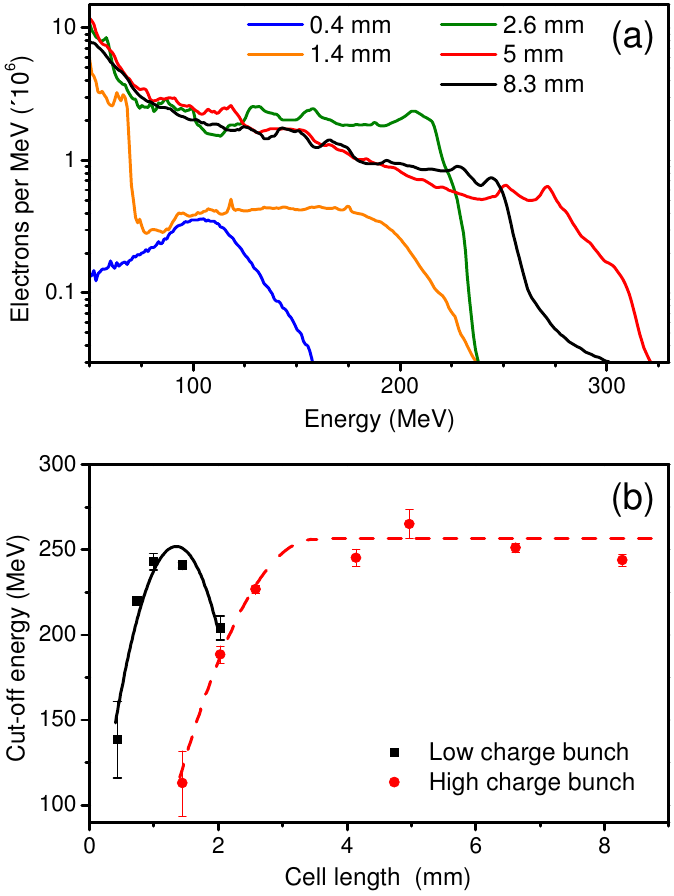}
	\caption{\textbf{Electron spectra for $\bm{n_e\approx 1.1\times 10^{19}}$~cm$\bm{^{-3}}$}. (a) Typical single shot spectra for different cell lengths. (b) Mean 5\% cut-off energy as a function of the cell length. The error bar is the standard error of the mean. The solid line is a parabolic fit, the dashed line is a visual guide. The peak accelerating electric field is $E_z^{max}=340\pm65$ GV/m for the first bunch and $E_z^{max}=185\pm40$ GV/m for the second bunch.}
	\label{fig:fig3}
\end{figure}

The analysis of the electron spectra clearly demonstrates the sequential injection of two electron bunches, with different charge yields and injection positions, but it does not give much insight on the injection mechanisms. More information can be obtained by analyzing the betatron X-ray radiation which arises from the transverse oscillations of accelerated electrons in the wake~\cite{PRL2004Kiselev,PRL2004Rousse}. The angular profile of the X-ray emission depends on the electron distribution in the transverse phase space, and as such can reveal important information on the transverse properties of the electron bunch~\cite{PRL2006TaPhuoc}. X-ray beam angular profiles, presented in Fig.~4, show that the betatron radiation produced by the two bunches has very different features. The X-ray beams radiated by the high charge bunch (Fig.~4e-h) are more intense, exhibit a lack of simple symmetry, off-center intensity peaks, flattened transverse profiles and are constantly changing in shape. In contrast, the low charge bunch emits very stable elliptical X-ray beams (Fig.~4a-d) with gaussian profile and major axis along the laser polarization (see Fig.~4i). 
The observed difference in intensity are essentially due to the variation of the electron bunch charge. 

\begin{figure}
\includegraphics[width=9 cm]{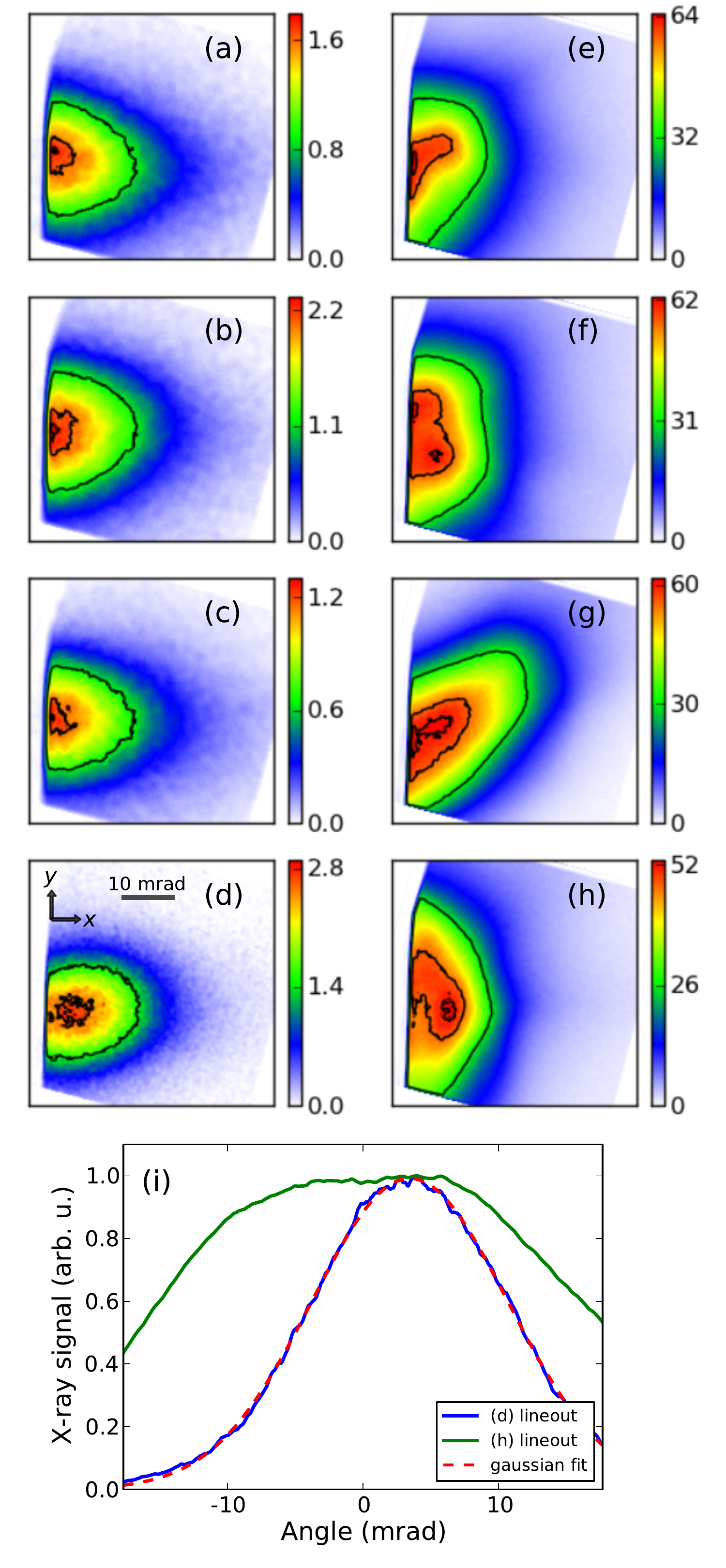}
\caption{\textbf{X-ray beam angular profiles for $\bm{n_e\approx1.1\times10^{19}}$~cm$\bm{^{-3}}$.} Typical angular profiles emitted by the low charge bunch for $L\approx 1.2$ mm (a)-(d) and the high charge bunch for $L\approx4$ mm (e)-(h). The field of view in each image is 44 mrad $\times$ 44 mrad and the color scale gives the number of counts recorded by the X-ray CCD, divided by $1000$. Photon noise is reduced by a mean filter, and contour lines at 50\%, 90\% and 98\% are shown for each image. Transverse axis $\hat{x}$ and $\hat{y}$ are defined in (d), where $\hat{x}$ is the laser polarization direction. (i) Lineout in the $\hat{y}$ direction (integrated over 2.25 mrad in the $\hat{x}$ direction) of image (d) and (h), and a gaussian fit of the (d) lineout. }
\label{fig:fig4}
\end{figure}

\begin{figure}
\includegraphics[width=9cm]{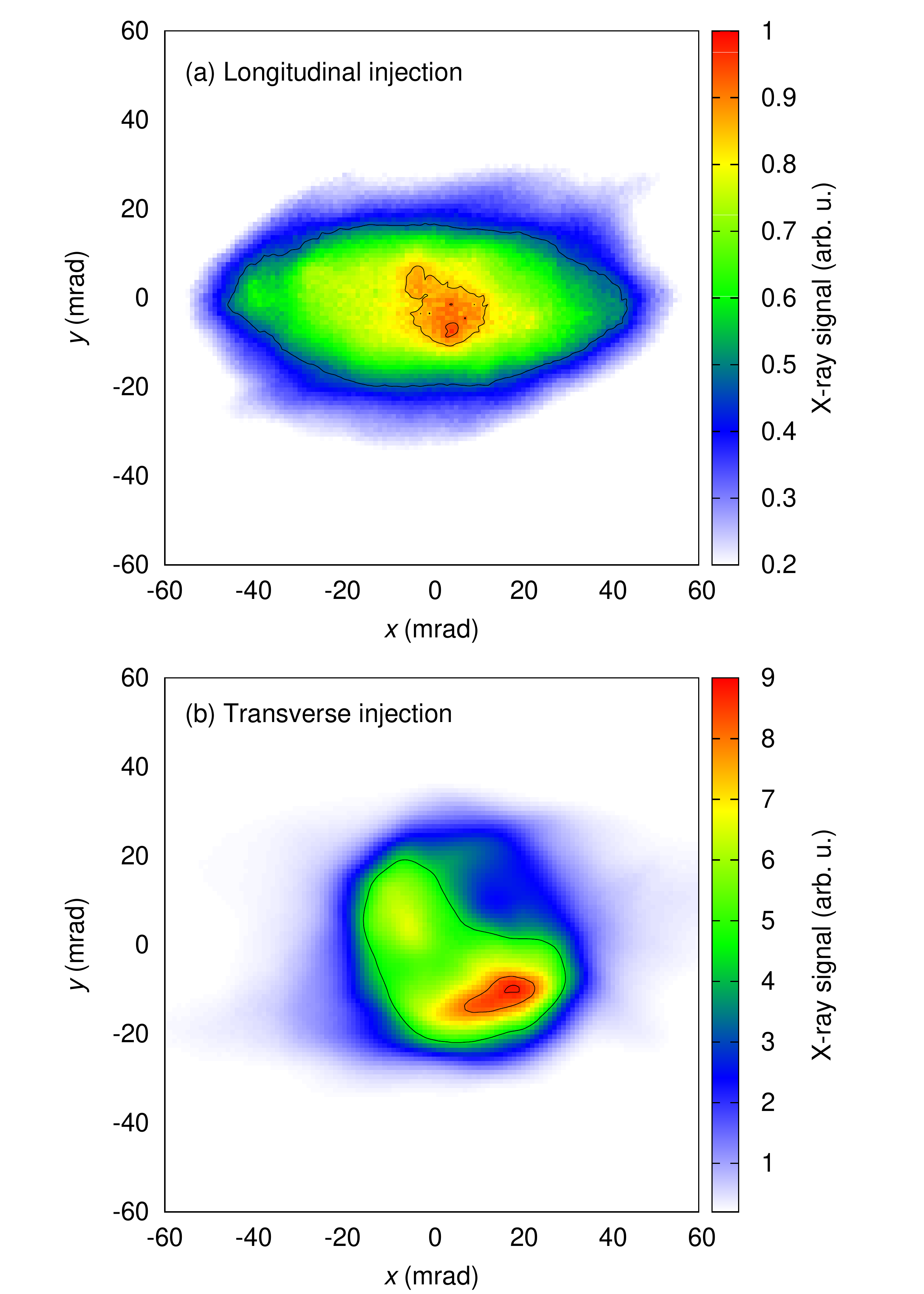}
\caption{\textbf{Calculated X-ray beam profiles.} The X-ray beam profiles are calculated from electron trajectories  coming from PIC simulation, for the first (a) and the second (b) bunches shown in Fig.~2. Contour lines at 50\%, 90\% and 98\% are shown.}
\label{fig:fig5}
\end{figure}

The asymmetry and shot-to-shot fluctuations of the X-ray beam profiles radiated by the high charge bunch suggest that its transverse distribution is also asymmetric and fluctuating, as expected for transverse injection (see Fig.~2d). To confirm that the high charge bunch in the experiment was actually trapped by transverse self-injection, we computed the X-ray emission radiated by the first and second injected bunches in Fig.~2. The results presented in Fig.~5 show that the X-ray beam emitted by the longitudinally injected bunch (first bunch) is elliptic and gaussian, just as the X-ray beams emitted by the low charge bunch in Fig.~4a-d. The ellipse major axis is along the laser polarization in both cases. The beam profile ellipticity  seems larger in simulation, suggesting that the electron bunch interacts more with the laser pulse in the simulation than in the experiment, probably because of slightly different conditions. The X-ray beam emitted by transversely injected electrons in Fig.~5b is asymmetric, presents off-center intensity peaks and a somewhat flattened profile. Theses features are similar to that of the experimental profiles in Fig.~4e-h, confirming that the electrons emitting these X-ray beams were trapped by transverse self-injection. 

In summary, we observed in both experiment and simulation a double injection. The first injection is due to longitudinal self-injection. It occurs very early in the laser propagation and leads to a charge of a few pC. The electron distribution is symmetric and stable shot-to-shot. The second injection is due to transverse self-injection. It occurs after significant laser self-focusing and leads to a charge of about 100 pC. The electron distribution in this case is asymmetric and changes shot-to-shot, because it depends strongly on the laser intensity profile at the injection time. 

\begin{figure}
	\includegraphics[width=9cm]{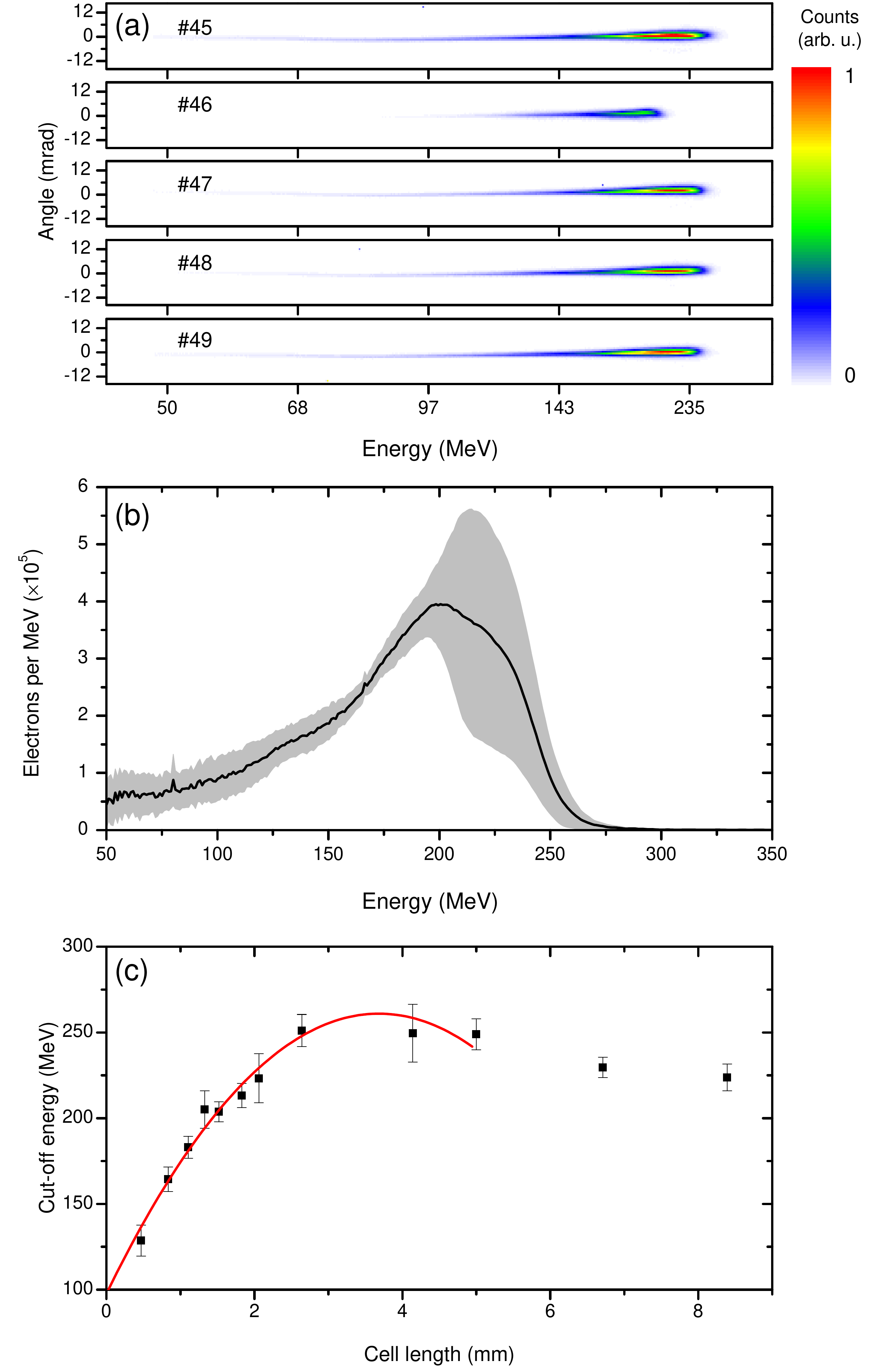}
	\caption{\textbf{Electron spectra for $\bm{n_e\approx7\times10^{18}}$~cm$\bm{^{-3}}$.}  (a) Five consecutive raw electron spectra, for  $L=2.7$~mm. (b) Mean deconvoluted spectrum. The gray area indicates the standard deviation band. (c) Mean 5\% cut-off energy as a function of the cell length. The error bar is the standard error of the mean. The parabolic fit for $L<5$ mm (red line) leads to $E_z^{max}=110\pm30$ GV/m.}
	\label{fig:fig6}
\end{figure}

\smallskip
\noindent
\textbf{A stable and tunable accelerator.}
Figure~4 showed that longitudinal injection yields much more stable electron beam properties than transverse injection. These properties can be further improved by reducing the electron density. 
 Experimental electron spectra from five consecutive shots, obtained for $n_e\approx 7\times 10^{18}$~cm$^{-3}$ and $L=2.7$ mm, are presented in Fig.~6a-b. They illustrate the good stability of the accelerator. For this density, only longitudinal injection is observed,  probably because laser self-focusing is not strong enough to get transverse injection. The fluctuations of the peak energy are  $4\%$ RMS, which is of the same order of magnitude as the laser intensity fluctuations. A similar stability is observed for all cell lengths. The measured beam charge is about 5 pC for all cell lengths, with shot-to-shot variations of  $25\%$ RMS. The probability of injection is $100\%$.   The divergence of the electron beam is smaller than 2 mrad full width at half maximum (FWHM), limited by the resolution of the detector and the energy spread is in the range 50-70 MeV FWHM.

X-ray measurements in Fig.~7 show stable and close-to-gaussian X-ray beam profiles, confirming the longitudinal self-injection mechanism for electron trapping. Compared to $n_e\approx 1.1\times10^{19}$ cm$^{-3}$, observed X-ray divergences are reduced down to 10 mrad FWHM. This indicates that the electron beam has a very small transverse emittance. The latter can be estimated as follows. The gaussian X-ray beam profile is consistent with a Maxwell-Boltzmann electron beam distribution in the transverse phase space. In this  case, the spatial variances $\sigma_a$ with $(a=x,y)$ and the angular variances $\sigma'_a$ of the electron beam are linked by $\sigma_a=\sigma'_{a}(2 \gamma / \alpha)^{1/2}k_p^{-1}$ with $\alpha$ the  normalized amplitude of the transverse focusing force inside the plasma cavity ($\alpha\leq1$ with $\alpha=1$ for a fully evacuated cavity). The normalized transverse emittance in absence of spatial-angular correlations is $\epsilon_{an} = \gamma \sigma_a\sigma'_{a}$. Considering that our experiment works close to full blow-out with $\alpha\in[0.5,1]$, we estimate $\epsilon_n\in[0.57,0.81]$ $\pi$.mm.mrad,  which is comparable to the smallest emittances measured in laser-plasma accelerators~\cite{PRL2010Brunetti,2012PRL_Plateau,weingartner2012}.
 
\begin{figure}
\includegraphics[width=9cm]{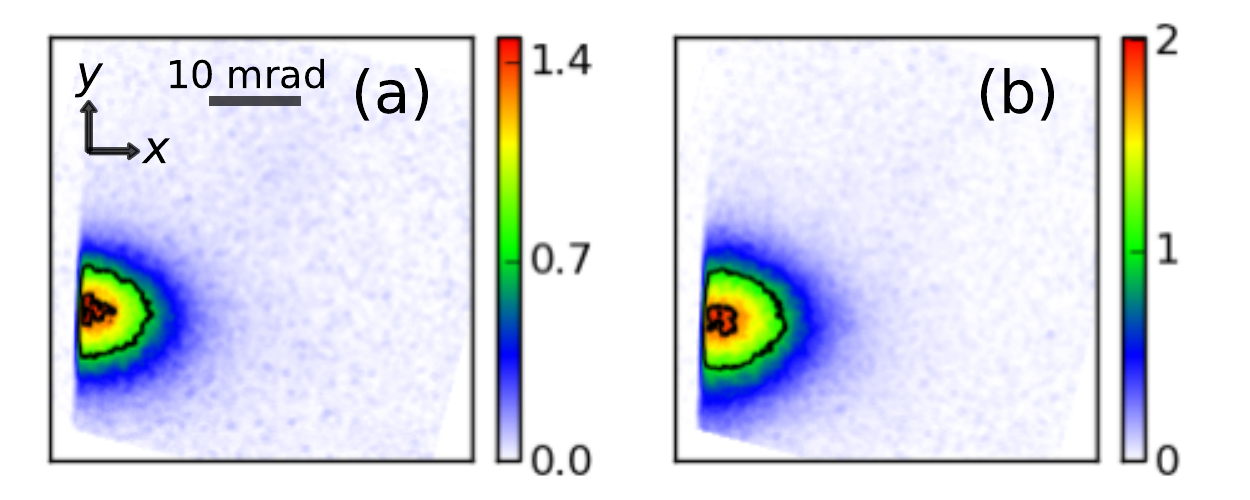}
\caption{\textbf{X-ray beam properties for $\bm{n_e\approx7\times10^{18}}$~cm$\bm{^{-3}}$.} (a)-(b) Two typical X-ray beam angular profiles past the dephasing length (scales and image processing are identical to Fig.~4). }
\label{fig:fig7}
\end{figure}

Reducing the density should in principle increase the peak electron energy.  Yet, Figs.~3b and~6c show that the measured peak energy is almost the same for both densities. The accelerating fields $E_z$ in both cases can be estimated from the parabolic fits in Figs.~3b and~6c. According to Lu's model~\cite{PRSTAB2007Lu}, the longitudinal electric field  $E_z$ should vary as $(a_0 n_e )^{1/2}$. Therefore, for a constant $a_0$, $E_z$ should  be divided by 1.3 when $n_e$ is divided by $1.6$.  In contrast, our measurements reveal that $E_z$ is divided by $3.1$. This strong decrease of $E_z$ is the main reason for the saturation of the peak electron energy at low density.  Several effects may explain the decrease of $E_z$.  For instance, the laser intensity reached after laser pulse self-focusing and self-steepening,  and the residual electron density in the plasma cavity can depend strongly on the plasma density.   Departure from a perfect bubble, $a_0$ dynamics and temporal evolution of the wakefield could also contribute to explain the large variations of $E_z$.
 In any case, the results show that the physics is more complex than suggested by simple scaling laws.

\bigskip
\noindent
{\textbf{Discussion}

\noindent
To improve the performance of laser-plasma accelerators and develop their applications, a precise understanding of the physics of these accelerators is mandatory, and  increasing efforts have been devoted toward this goal~\cite{PRL2006Hsieh,matlis2006,NatPhys2011Buck,corde_mapping}. Here we used a gas cell with an adaptable length to directly observe the injection, acceleration and dephasing of electrons. 
We demonstrated that electrons can be trapped either via longitudinal or transverse self-injection, and that these two types of injection lead to energetic electrons having very different features. Transverse injection is well suited for applications which require a high charge ($\gtrsim 100$ pC) but can cope with little stability, a large energy spread, and a mediocre emittance. In contrast, longitudinal injection is ideal for applications in which a good stability and a low emittance are essential.

\bigskip
\noindent
{\textbf{Methods}}

{\small
\noindent
\textbf{Particle-in-cell simulations.} 
3D fully electromagnetic particle-in-cell simulations were
performed with the numerical code Calder Circ, described in
Ref. \cite{calder-circ}, which uses a Fourier decomposition of the
electromagnetic fields in the azimuthal direction with respect to the laser propagation direction. Seven Fourier modes
($m=0$ to $m=6$) were included in the simulations. The laser wavelength is $\lambda_0=800$ nm, the normalized laser peak amplitude $a_0=2$, the duration is 30 fs FWHM and the spatial intensity profile is taken as  the projection over the modes of the experimental laser spot (shown in Fig. 2a). The mesh  consists of 5000 (in $z$) $\times$ 600 (in $r$) cells. The cell size is
$0.096 \lambda_0/2 \pi$ (in $z$) $\times$ $1.5  \lambda_0/2 \pi$ (in
$r$) and the time step is $0.066 \lambda_0/2 \pi c$. Positive ions are considered in the simulations as a uniform frozen background. The number of electron macroparticles per cell is 750. The longitudinal plasma profile consists of a 65 $\mu$m long linear ramp followed by a 1.5 mm long plateau with density $n_0=0.006 n_c=9.6 \times 10^{18}$ cm$^{-3}$.

\noindent
\textbf{Laser system.} 
The experiment was conducted at Laboratoire d'Optique Appliqu\'ee with the ``Salle Jaune'' Ti:Sa laser system, which delivers 1 Joule on target with a full width at half maximum (FWHM) pulse duration of 35 fs and a linear polarization. The laser pulse was focused at the entrance of the gas cell with a 70-cm-focal-length off-axis parabola, to a FWHM focal spot size of 20~$\mu$m. From the measured intensity distribution in the focal plane, the peak intensity was estimated to be $4\times10^{18}$ W.cm$^{-2}$, corresponding to a normalized amplitude of $a_0=1.4$.

\noindent
\textbf{Gas cell.} 
The target was a cell filled with neutral helium gas and whose back face can be translated continuously under vacuum, allowing to change its length in the laser propagation direction. The front face of the cell is fixed in order to keep the same focal plane position with respect to the target entrance. The front and back faces had a 500~$\mu$m diameter hole, respectively for the entrance of the laser pulse in the gas cell and for the exit of the electron and X-ray beams. A characterization of the gas cell atomic density realized  by interferometry showed a flat longitudinal density profile and yielded the calibration between backing pressure and electron density.

\noindent
\textbf{Electron and X-ray beam diagnostics.} 
The electron spectrometer consisted of a permanent bending magnet (1.1~T over 10~cm) which deflects electrons depending on their energy, and a Lanex phosphor screen to convert a fraction of the electron energy into 546 nm light imaged by a 16 bits visible CCD camera. X-ray beams were measured with an X-ray CCD camera with $2048\times2048$ pixels of size $13.5\times13.5$ $\mu$m$^2$), set at 60 cm from the gas cell and protected from the laser light by a 12 $\mu$m Al filter.
}

\bigskip
\noindent
{

\bigskip
\noindent
{\textbf{Acknowledgments}}

\noindent
This work was supported by the European Research Council (PARIS ERC, Contract No. 226424) and EuCARD/ANAC, EC FP7 (Contract No. 227579). It was partly performed using high-performance computing resources of GENCI (Grant No. 2012-056304). 

\bigskip
\noindent
{\textbf{Authors contributions}}

\noindent
S.C., C.T., G.L. and K.T.P conceived and realized the experiment. S.C. and C.T.  analyzed the data. A.L and X.D performed the simulations. S.C., C.T., A.L., X.D., R.L. and V.M. discussed the results. S.C., C.T., A.L. and R.L wrote the paper. G.L. and D.D. designed the gas cell. V.M and A.R.  provided overall guidance to the project.

\bigskip
\noindent
{\textbf{Correspondence}}

\noindent
Correspondence should be addressed to C.T.~(email: cedric.thaury@ensta-paristech.fr).

\end{document}